\documentstyle[aps,epsf,aps10]{revtex}
\begin{document}

\def\slash#1{#1 \hskip -0.5em / } 
\def\beq{\begin{equation}}
\def\eeq{\end{equation}}
\def\beqy{\begin{eqnarray}}
\def\eeqy{\end{eqnarray}}

\title{On the Monopole Model Of Form Factors for Heavy Meson Decays}
\author{W. Roberts}
\address{Department of Physics, Old Dominion University\\
Norfolk, VA
23529, USA\\
and\\
 Thomas Jefferson National Accelerator Facility,
12000 Jefferson Avenue, Newport News, VA 23606, USA.}
\maketitle
\begin{abstract}
A pole model for the form factors describing heavy to light meson transitions is
constructed, and the results compared with the predictions of the heavy quark
effective theory (HQET). While most of the HQET-predicted relationships among
form factors hold at the maximum value of $q^2$, many are violated, suggesting
that the simplest version of the pole model of meson form factors should be used with caution,
especially when combined with HQET.

\vspace{5mm}
 
\flushright{JLAB-THY-02-13}
\end{abstract}
\thispagestyle{empty}
\setcounter{page}{1}

\section{Introduction and Motivation}

Over the past decade or so, the heavy quark effective theory (HQET) \cite{hqet} has proven to be a very useful tool for studying many
aspects of the phenomenology of heavy quarks and the hadrons containing them. The predictions of this effective
field theory have been used for `model independent' extractions of $V_{cb}$, and similar extractions are being attempted for
$V_{ub}$. Both inclusive and exclusive semileptonic decays have seen much activity in this field, but there have
also been many applications to rare radiative decays and leptonic decays and, through various factorization
approximations, weak nonleptonic decays, as well as rare ones, have also been treated, with some success.

One set of processes in which the predictions of the HQET appeared to fail was in the rare decays $B\to
K^*\gamma$, $B\to K\psi^{(\prime)}$ and $B\to K^*\psi^{(\prime)}$. In particular, the predictions for the degree of longitudinal polarization of the vector
mesons
in the last process was found by many authors to be significantly different from the measured value. In treating these processes, the form factors
extracted from $D\to K^{(*)}$ semileptonic decays were taken and extrapolated to the much larger kinematic regime
of $B$ decays, and together with factorization, were used to describe the $B\to K^{(*)}\psi^{(\prime)}$
processes. The failure of the HQET predictions has led to the suggestion that there are non-factorizable contributions 
to the decay matrix element \cite{nonfact,orsay}.

It is possible that this conclusion is unnecessary, or that the size of the `non-factorizable contributions' are smaller than estimated.
A possible source of the discrepancy between theory and experiment could lie in the forms chosen for the
semileptonic form factors that describe the $D\to K^{(*)}$ semileptonic decays. Since these form factors formed
the crucial basis of all that followed, any `errors' in them would propagate into the later predictions. Indeed,
using the same assumptions of HQET, with factorization, we found a number of scenarios that successfully
described all of the then available data, without the need for invoking other mechanisms or non-factorizable contributions, provided that
the forms of the $D\to K^{(*)}$ semileptonic form factors were parametrized in a manner different from that
chosen by the experimentalists \cite{robertsledroit}.

To illustrate the problem, let's look at the decays $D\to K^{(*)}\ell\nu$, both within the context of HQET, and
otherwise. The weak matrix elements for these decay are parametrized in terms of six form factors as
\beqy
\left<K(p^\prime)\left|\bar{s}\gamma_\mu c\right|D(p)\right>&=&f_+\left(p+p^\prime\right)_\mu+
f_-\left(p-p^\prime\right)_\mu,\nonumber \\
\left<K^*(p^\prime,\varepsilon)\left|\bar{s}\gamma_\mu c\right|D(p)\right>&=&
ig\epsilon_{\mu\nu\alpha\beta}\varepsilon^{*\nu}\left(p+p^\prime\right)^\alpha \left(p-p^\prime\right)^\beta,\nonumber \\
\left<K^*(p^\prime,\varepsilon)\left|\bar{s}\gamma_\mu\gamma_5 c\right|D(p)\right>&=&f\varepsilon^*_\mu+
\varepsilon^*\cdot p\left[a_+\left(p+p^\prime\right)_\mu+
a_-\left(p-p^\prime\right)_\mu\right].
\eeqy
In the context of HQET, the $D$ meson is represented as \cite{falkstates}
\beq
\left|{\cal D}(v)\right>=-\sqrt{\frac{m_D}{2}}\frac{1+\slash{v}}{2}\gamma_5\equiv {\cal M}_D(v),
\eeq
where $v=p/m_D$, and the matrix elements are obtained as, at leading order in HQET \cite{mannelroberts}
\beqy
\left<K(p^\prime)\left|\bar{s}\Gamma h_v^{(c)}\right|{\cal D}(v)\right>&=&{\rm Tr}
\left[\gamma_5\left(\xi_1+\slash{p}\xi_2\right)\Gamma{\cal M}_D(v)\right],\nonumber \\
\left<K^*(p^\prime,\varepsilon)\left|\bar{s}\Gamma h_v^{(c)}\right|{\cal D}(v)\right>&=&{\rm Tr}
\left[\left\{\left(\xi_3+\slash{p}\xi_4\right)\varepsilon^*\cdot v+\slash{\varepsilon}^*\left(\xi_5+\slash{p}\xi_6\right)
\right\}\Gamma{\cal M}_D(v)\right].
\eeqy
The functions $\xi_i$ are independent of the mass of the heavy quark, and would be the same if the initial meson
were a $B$ or $T$ meson instead.

Taking the traces for the specific cases above gives
\beqy
\left<K(p^\prime)\left|\bar{s}\gamma_\mu c\right|D(p)\right>&=&-\sqrt{2 m_D}\left(\xi_1 v_\mu-\xi_2
p^\prime_\mu\right),\nonumber \\
\left<K^*(p^\prime,\varepsilon)\left|\bar{s}\gamma_\mu c\right|D(p)\right>&=&-i\sqrt{2 m_D}\xi_6
\epsilon_{\mu\nu\alpha\beta}\varepsilon^{*\nu}p^{\prime\alpha}v^\beta,\nonumber \\
\left<K^*(p^\prime,\varepsilon)\left|\bar{s}\gamma_\mu\gamma_5 c\right|D(p)\right>&=&\sqrt{2m_D}\left[\xi_3
\varepsilon^*\cdot v v_\mu+\left(\xi_6-\xi_4\right)\varepsilon^*\cdot v p^\prime_\mu-\left(\xi_5+v\cdot
p^\prime\xi_6\right)\varepsilon^*_\mu\right].
\eeqy
Comparing the two sets of matrix elements leads to
\beqy \label{s_light}
f_+&=&-\frac{1}{\sqrt{2m_D}}\left(m_D\xi_2+\xi_1\right),\nonumber\\
f_-&=&\frac{1}{\sqrt{2m_D}}\left(\xi_1-m_D\xi_2\right),\nonumber\\
g&=&\frac{1}{\sqrt{2m_D}}\xi_6,\nonumber\\
f&=&-\sqrt{2m_D}\left(\xi_5+v\cdot p^\prime\xi_6\right),\nonumber\\
a_+&=&\frac{1}{\sqrt{2m_D^3}}\left[\xi_3+m_D\left(\xi_6-\xi_4\right)\right],\nonumber\\
a_-&=&\frac{1}{\sqrt{2m_D^3}}\left[\xi_3-m_D\left(\xi_6-\xi_4\right)\right].
\eeqy

If, on the other hand, the strange quark were to be treated as heavy, then all of the form factors could be
written in terms of a single Isgur-Wise function, $\xi(w)$, as
\beqy \label{s_heavy}
f_+&=&-\frac{1}{4\sqrt{m_Dm_K}}\left(m_D+m_K\right)\xi(w),\nonumber\\
f_-&=&-\frac{1}{4\sqrt{m_Dm_K}}\left(m_D-m_K\right)\xi(w),\nonumber\\
g&=&\frac{1}{4\sqrt{m_Dm_K}}\xi(w),\nonumber\\
f&=&-\frac{\sqrt{m_Dm_K}}{2}\left(1+w\right)\xi(w),\nonumber\\
a_+&=&-a_-=\frac{1}{4\sqrt{m_Dm_K}}\xi(w),\nonumber\\
\eeqy
where $w=v\cdot v^\prime$ and $v^\prime=p^\prime/m_K^{(*)}$. These relationships are obtainable from the previous ones by noting that, by comparison with the
Falk representations of states, in the limit $m_s\to\infty$,
\beqy
\xi_1&\to&\frac{1}{2\sqrt{2}}\sqrt{m_K} \xi,\nonumber\\
\xi_2&\to&\frac{1}{2\sqrt{2}}\frac{1}{\sqrt{m_K}}\xi,\nonumber\\
\xi_3&\to&0,\nonumber\\
\xi_4&\to&0,\nonumber\\
\xi_5&\to&\frac{1}{2\sqrt{2}}\sqrt{m_K} \xi,\nonumber\\
\xi_6&\to&\frac{1}{2\sqrt{2}}\frac{1}{\sqrt{m_K}}\xi.
\eeqy

It has become customary to parametrize these form factors as monopole forms and, in fact, this is how the available data
in $D\to K^{(*)}$ semileptonic decays have been treated \cite{cleo}. However, from eqns. (\ref{s_light}) and (\ref{s_heavy}),
it is clear that using simple monopole forms for {\it all} of the form factors is contradictory to the
predictions of HQET. In particular, the form factor $f$ can not be a simple monopole. Indeed, if the Isgur-Wise
function of eqn. (\ref{s_heavy}) is chosen to be a simple monopole, then the $q^2$ dependence in $f$ is not that
of a monopole. While the differences in parametrization may have insignificant impact on the analysis of the
$D\to K^{(*)}$ semileptonic decays, the consequences can be very significant for $B\to K^{(*)}$ decays. Note that 
Aleksan {\it et al.} \cite{orsay} have pointed out that $f$ should increase more slowly with $q^2$ than $f_+$, $g$ or $a_+$, in order to
accomodate the nonleptonic and rare data. 

The question of the parametrization of these form factors is also important for the semileptonic decays of $B$ mesons to
light ones like $\rho$ and $\pi$, as these processes provide one means of obtaining the CKM matrix element $V_{ub}$. Some
information on the $B\to\pi$ and $B\to\rho$ transition form factors can be obtained by applying HQET to the $D\to\pi$ and
$D\to\rho$ semileptonic transitions. It has been proposed to extract these latter form factors with great precision at 
CLEO-c, and these could then be used in the extraction of $V_{ub}$ from the corresponding $B$ decays. However, the value
obtained for this CKM matrix element could be strongly dependent on the parametrization of the form factors. 

One question that arises here is the following. Is the monopole form valid, even for the Isgur-Wise function $\xi(w)$, or
for the $\xi_i$? Is this a `model' or an ansatz for the form factors that is consistent with HQET? To examine this, we have
constructed a monopole model of form factors, and compared the predictions with those of HQET. The monopole model
is constructed within the framework of HQET. In the next section, we discuss a number of relations among form
factors that arise in HQET. For concreteness, we examine the `decays' of $D$ and $D^*$ mesons to $K$ and $K^*$
mesons, through the scalar, pseudoscalar, vector, axial vector and tensor currents. In section III, we present
the pole model we use, and point out some of the key results obtained there. In section IV we compare the
predictions of HQET with those of the pole model, and in section V we present our conclusions.

\section{Form Factors}

In this section, we discuss the transitions $D^{(*)}\to K^{(*)}$ through the
scalar, pseudoscalar, vector, axial vector and tensor currents. We first give
the general Lorentz structure of the matrix elements, then examine the
structures that arise in HQET, if the strange quark is treated as a light quark.
In this case, the matrix elements are written in terms of the 6 functions
$\xi_i$, introduced previously.

The 20 matrix elements of interest are written in terms of form factors as
\beqy\label{lorentza}
\left<K(p^\prime)\left|\bar{s} c\right|D(p)\right>&=&S_1(q^2),\nonumber\\
\left<K^*(p^\prime,\varepsilon^\prime)\left|\bar{s} c\right|D(p)\right>&=&
\left<K(p^\prime)\left|\bar{s}c\right|D^*(p,\varepsilon)\right>=0,\nonumber\\
\left<K^*(p^\prime,\varepsilon^\prime)\left|\bar{s}
c\right|D^*(p,\varepsilon)\right>&=&S_2(q^2)\varepsilon\cdot\varepsilon^{\prime*}+S_3(q^2) \varepsilon\cdot
p^\prime\varepsilon^{\prime*}\cdot p,\nonumber\\
\left<K(p^\prime)\left|\bar{s}\gamma_5 c\right|D(p)\right>&=&0,\nonumber\\
\left<K^*(p^\prime,\varepsilon^\prime)\left|\bar{s}\gamma_5 c\right|D(p)\right>&=&r_1(q^2)\varepsilon^{\prime*}\cdot p,\nonumber\\
\left<K(p^\prime)\left|\bar{s}\gamma_5 c\right|D^*(p,\varepsilon)\right>&=&r_2(q^2)\varepsilon\cdot p^\prime,\nonumber\\
\left<K^*(p^\prime,\varepsilon^\prime)\left|\bar{s}\gamma_5
c\right|D^*(p,\varepsilon)\right>&=&ir_3(q^2)\epsilon_{\mu\nu\alpha\beta}\varepsilon^\mu\varepsilon^{\prime\nu}
\left(p+p^\prime\right)^\alpha\left(p-p^\prime\right)^\beta=
-2ir_3(q^2)\epsilon_{\mu\nu\alpha\beta}\varepsilon^\mu\varepsilon^{\prime\nu}p^\alpha p^{\prime\beta},\nonumber\\
\left<K(p^\prime)\left|\bar{s}\gamma_\mu c\right|D(p)\right>&=&f_+(q^2)\left(p+p^\prime\right)_\mu+
f_-(q^2)\left(p-p^\prime\right)_\mu,\nonumber\\
\left<K^*(p^\prime,\varepsilon^\prime)\left|\bar{s}\gamma_\mu c\right|D(p)\right>&=&
ig(q^2)\epsilon_{\mu\nu\alpha\beta}\varepsilon^{\prime*\nu}\left(p+p^\prime\right)^\alpha \left(p-p^\prime\right)^\beta,\nonumber\\
\left<K(p^\prime)\left|\bar{s}\gamma_\mu c\right|D^*(p,\varepsilon)\right>&=&
ig^\prime(q^2)\epsilon_{\mu\nu\alpha\beta}\varepsilon^{*\nu}\left(p+p^\prime\right)^\alpha \left(p-p^\prime\right)^\beta,\nonumber\\
\left<K^*(p^\prime,\varepsilon^\prime)\left|\bar{s}\gamma_\mu c\right|D^*(p,\varepsilon)\right>&=&v_1(q^2)\varepsilon\cdot
p^\prime\varepsilon^{\prime*}_\mu+v_2(q^2)\varepsilon^{\prime*}\cdot p\varepsilon_\mu\nonumber\\
&+&\varepsilon\cdot p^\prime\varepsilon^{\prime*}\cdot p\left[v_+(q^2)\left(p+p^\prime\right)_\mu+v_-(q^2)\left(p-p^\prime\right)_\mu\right]\nonumber\\
&+&\varepsilon\cdot\varepsilon^{\prime*}\left[v_+^\prime(q^2)\left(p+p^\prime\right)_\mu+v_-^\prime(q^2)\left(p-p^\prime\right)_\mu\right]\nonumber\\
\left<K(p^\prime)\left|\bar{s}\gamma_\mu\gamma_5 c\right|D(p)\right>&=&0,\nonumber\\
\left<K^*(p^\prime,\varepsilon^\prime)\left|\bar{s}\gamma_\mu\gamma_5 c\right|D(p)\right>&=&f(q^2)\varepsilon^{\prime*}_\mu+
\varepsilon^{\prime*}\cdot p\left[a_+(q^2)\left(p+p^\prime\right)_\mu+
a_-(q^2)\left(p-p^\prime\right)_\mu\right]\nonumber\\
\left<K(p^\prime)\left|\bar{s}\gamma_\mu\gamma_5 c\right|D^*(p,\varepsilon)\right>&=&f^\prime(q^2)\varepsilon_\mu+
\varepsilon\cdot p^\prime\left[a_+^\prime(q^2)\left(p+p^\prime\right)_\mu+
a_-^\prime(q^2)\left(p-p^\prime\right)_\mu\right]\nonumber\\
\left<K^*(p^\prime,\varepsilon^\prime)\left|\bar{s}\gamma_\mu\gamma_5 c\right|D^*(p,\varepsilon)\right>&=&i\epsilon_{\mu\nu\alpha\beta}
\left\{\left(p+p^\prime\right)^\alpha\left(p-p^\prime\right)^\beta\left[h_1(q^2)\varepsilon\cdot p^\prime\varepsilon^{\prime*\nu}+
h_2(q^2)\varepsilon^{\prime*}\cdot p\varepsilon^\nu\right]\right.\nonumber\\
&+&\left.\varepsilon^\nu\varepsilon^{\prime*\alpha}\left[h_+(q^2)\left(p+p^\prime\right)_\beta+h_-(q^2)\left(p-p^\prime\right)_\beta\right]\right\}\nonumber\\
\left<K(p^\prime)\left|\bar{s}\sigma_{\mu\nu} c\right|D(p)\right>&=&is(q^2)\left[\left(p+p^\prime\right)_\mu\left(p-p^\prime\right)_\nu-
\left(p-p^\prime\right)_\mu\left(p+p^\prime\right)_\nu\right]\nonumber\\
\left<K^*(p^\prime,\varepsilon^\prime)\left|\bar{s}\sigma_{\mu\nu} c\right|D(p)\right>&=&\epsilon_{\mu\nu\alpha\beta}\left\{h(q^2)\varepsilon^{\prime*}\cdot p
\left(p+p^\prime\right)^\alpha\left(p-p^\prime\right)^\beta+\varepsilon^{\prime*\alpha}\left[g_+(q^2)\left(p+p^\prime\right)^\beta+
g_-(q^2)\left(p-p^\prime\right)^\beta\right]\right\},\nonumber\\
\left<K(p^\prime)\left|\bar{s}\sigma_{\mu\nu}c\right|D^*(p,\varepsilon)\right>&=&\epsilon_{\mu\nu\alpha\beta}\left\{h^\prime(q^2)\varepsilon\cdot p^\prime
\left(p+p^\prime\right)^\alpha\left(p-p^\prime\right)^\beta+\varepsilon^\alpha\left[g_+^\prime(q^2)\left(p+p^\prime\right)^\beta+
g_-^\prime(q^2)\left(p-p^\prime\right)^\beta\right]\right\},\nonumber\\
\left<K^*(p^\prime,\varepsilon^\prime)\left|\bar{s}\sigma_{\mu\nu}c\right|D^*(p,\varepsilon)\right>&=&it_0(q^2)
\left(\varepsilon_\mu\varepsilon^{\prime*}_\nu-\varepsilon_\nu\varepsilon^{\prime*}_\mu\right)\nonumber\\
&+&
i\left[\left(p+p^\prime\right)_\mu\left(p-p^\prime\right)_\nu-\left(p+p^\prime\right)_\nu\left(p-p^\prime\right)_\mu\right]
\left(t_1(q^2)\varepsilon\cdot\varepsilon^{\prime*}+t_2(q^2)\varepsilon\cdot p^\prime\varepsilon^{\prime*}\cdot
p\right)\nonumber \\
&+&i\varepsilon^{\prime*}\cdot p\left\{t_+(q^2)\left[\varepsilon_\mu\left(p+p^\prime\right)_\nu-
\varepsilon_\nu\left(p+p^\prime\right)_\mu\right]+t_-(q^2)\left[\varepsilon_\mu\left(p-p^\prime\right)_\nu-
\varepsilon_\nu\left(p-p^\prime\right)_\mu\right]\right\}\nonumber\\
&+&i\varepsilon\cdot p^\prime\left\{t_+^\prime(q^2)\left[\varepsilon_\mu^{\prime*}\left(p+p^\prime\right)_\nu-
\varepsilon_\nu^{\prime*}\left(p+p^\prime\right)_\mu\right]+t_-^\prime(q^2)\left[\varepsilon_\mu^{\prime*}\left(p-p^\prime\right)_\nu-
\varepsilon_\nu^{\prime*}\left(p-p^\prime\right)_\mu\right]\right\}.
\eeqy

These 20 matrix elements are thus expressed in terms of 40 {\it a priori}
independent form factors. 

In terms of the 6 functions $\xi_i$, at leading order in HQET, the matrix elements above are found to be
\beqy\label{heavya}
\left<K(p^\prime)\left|\bar{s} c\right|D(p)\right>&=&-\sqrt{2m_D}\left[v\cdot p^\prime\xi_2+\xi_1\right],\nonumber\\
\left<K^*(p^\prime,\varepsilon^\prime)\left|\bar{s} c\right|D^*(p,\varepsilon)\right>&=&
\sqrt{2m_D}\left[\varepsilon\cdot\varepsilon^{\prime*}\left(\xi_5+v\cdot p^\prime\xi_6\right)+\varepsilon\cdot
p^\prime\varepsilon^{\prime*}\cdot v\left(\xi_4-\xi_6\right)\right],\nonumber\\
\left<K^*(p^\prime,\varepsilon^\prime)\left|\bar{s}\gamma_5 c\right|D(p)\right>&=&\sqrt{2m_D}\varepsilon^{\prime*}\cdot v\left(v\cdot p^\prime
\xi_4+\xi_5-\xi_3\right),\nonumber\\
\left<K(p^\prime)\left|\bar{s}\gamma_5 c\right|D^*(p,\varepsilon)\right>&=&-\sqrt{2m_D}\varepsilon\cdot p^\prime\xi_2,\nonumber\\
\left<K^*(p^\prime,\varepsilon^\prime)\left|\bar{s}\gamma_5
c\right|D^*(p,\varepsilon)\right>&=&-i\sqrt{2m_D}\xi_6\epsilon_{\mu\nu\alpha\beta}\varepsilon^\mu\varepsilon^{\prime *\nu}
p^{\prime\alpha} v^\beta,\nonumber\\
\left<K(p^\prime)\left|\bar{s}\gamma_\mu c\right|D(p)\right>&=-&\sqrt{2m_D}\left(\xi_2 p^\prime_\mu+\xi_1 v_\mu\right),\nonumber\\
\left<K^*(p^\prime,\varepsilon^\prime)\left|\bar{s}\gamma_\mu c\right|D(p)\right>&=&i\sqrt{2m_D}
\xi_6\epsilon_{\nu\mu\alpha\beta}\varepsilon^{\prime*\nu}p^{\prime^\alpha}v^\beta,\nonumber\\
\left<K(p^\prime)\left|\bar{s}\gamma_\mu c\right|D^*(p,\varepsilon)\right>&=&i\sqrt{2m_D}\xi_2
\epsilon_{\nu\mu\alpha\beta}\varepsilon^{*\nu}p^{\prime\alpha}v^\beta,\nonumber\\
\left<K^*(p^\prime,\varepsilon^\prime)\left|\bar{s}\gamma_\mu c\right|D^*(p,\varepsilon)\right>&=&\sqrt{2m_D}\left[-\varepsilon\cdot
p^\prime\varepsilon^{\prime*}_\mu\xi_6+\varepsilon^{\prime*}\cdot v\varepsilon_\mu\left(\xi_3-v\cdot p^\prime\xi_4-\xi_5\right)\right.\nonumber\\
&+&\left.\varepsilon\cdot p^\prime\varepsilon^{\prime*}\cdot p\xi_4 v_\mu+
\varepsilon\cdot\varepsilon^{\prime*}\left(\xi_6 p^\prime_\mu+\xi_5 v_\mu\right)\right],\nonumber\\
\left<K^*(p^\prime,\varepsilon^\prime)\left|\bar{s}\gamma_\mu\gamma_5 c\right|D(p)\right>&=&\sqrt{2m_D}\left\{-\left(\xi_5+v\cdot
p^\prime\xi_6\right)\varepsilon^{\prime*}_\mu+
\varepsilon^{\prime*}\cdot v\left[\xi_3 v_\mu+\left(\xi_6-\xi_4\right)p^\prime_\mu\right]\right\},\nonumber\\
\left<K(p^\prime)\left|\bar{s}\gamma_\mu\gamma_5 c\right|D^*(p,\varepsilon)\right>&=&\sqrt{2m_D}
\left[-\left(\xi_1+v\cdot p^\prime \xi_2\right)\varepsilon_\mu+
\varepsilon\cdot p^\prime\xi_2 v_\mu\right],\nonumber\\
\left<K^*(p^\prime,\varepsilon^\prime)\left|\bar{s}\gamma_\mu\gamma_5 c\right|D^*(p,\varepsilon)\right>&=&-i\sqrt{2m_D}\epsilon_{\mu\nu\alpha\beta}
\left[\varepsilon^\alpha\varepsilon^{\prime*\beta}\left(\xi_6 p^\prime_\nu+\xi_5 v_\nu\right)+\xi_4\varepsilon^{\prime*}\cdot v\varepsilon^\nu
p^{\prime\alpha} v^\beta\right],\nonumber\\
\left<K(p^\prime)\left|\bar{s}\sigma_{\mu\nu} c\right|D(p)\right>&=&-i\sqrt{2m_D}\xi_2\left(p^\prime_\mu v_\nu-p^\prime_\nu v_\mu\right),\nonumber\\
\left<K^*(p^\prime,\varepsilon^\prime)\left|\bar{s}\sigma_{\mu\nu} c\right|D(p)\right>&=&\sqrt{2m_D}\epsilon_{\mu\nu\alpha\beta}
\left[\xi_4\varepsilon^{\prime*}\cdot v
p^{\prime\alpha}v^\beta+\varepsilon^{\prime*\alpha}\left(\xi_6p^{\prime\beta}+\xi_5v^\beta\right)\right],\nonumber\\
\left<K(p^\prime)\left|\bar{s}\sigma_{\mu\nu}c\right|D^*(p,\varepsilon)\right>&=&\sqrt{2m_D}\epsilon_{\mu\nu\alpha\beta}\varepsilon^\alpha\left(
\xi_1 v^\beta+\xi_2 p^{\prime\beta}\right),\nonumber\\
\left<K^*(p^\prime,\varepsilon^\prime)\left|\bar{s}\sigma_{\mu\nu}c\right|D^*(p,\varepsilon)\right>&=&i\sqrt{2m_D}\left\{\left(\xi_5+v\cdot 
p^\prime \xi_6\right)
\left(\varepsilon_\mu\varepsilon^{\prime*}_\nu-\varepsilon_\nu\varepsilon^{\prime*}_\mu\right)\right.\nonumber\\
&+&
\xi_6\left[-\left(v_\mu p^\prime_\nu-v_\nu p^\prime_\mu\right)\varepsilon\cdot\varepsilon^{\prime*}+\varepsilon\cdot p^\prime\left(
v_\mu\varepsilon_\nu^{\prime*}-v_\nu\varepsilon_\mu^{\prime*}\right)\right]\nonumber \\
&+&\left.\varepsilon^{\prime*}\cdot v\left[\left(\xi_6-\xi_4\right)\left(\varepsilon_\mu p^\prime_\nu-
\varepsilon_\nu p^\prime_\mu\right)+\xi_3\left(\varepsilon_\mu v_\nu-
\varepsilon_\nu v_\mu\right)\right]\right\}.
\eeqy

Because there are 40 form factors in eqns. (\ref{lorentza}) and only 6 functions
in eqns. (\ref{heavya}), a large number of relationships among the form factors
of eqns. (\ref{lorentza}) can be derived. Among these are
\beqy \label{relationships}
a_+^\prime-a_-^\prime&=&h^\prime=h_1=t_2=t_+^\prime-t_-^\prime=v_+-v_-=0,\nonumber\\
f_++f_-&=&-\left(g_+^\prime+g_-^\prime\right),\nonumber\\
f_+-f_-&=&=r_2=2m_Dg^\prime=-m_D\left(a_+^\prime+a_-^\prime\right)=2m_D
s=g_-^\prime-g_+^\prime,\nonumber\\
t_+^\prime+t_-^\prime&=&2r_3=2g=\frac{v_1}{m_D}=-\frac{v_+^\prime-v_-^\prime}{m_D}=
\frac{h_+-h_-}{m_D}=-\frac{g_+-g_-}{m_D}=-2t_1,\nonumber\\
a_++a_-&=&t_++t_-,\nonumber\\
v_++v_-&=&-2h_2=2h,\nonumber\\
v_+^\prime+v_-^\prime&=&-\left(h_++h_-\right)=g_++g_-,\nonumber\\
a_--a_+&=&=s_3=t_--t_+,\nonumber\\
s_1&=&f^\prime,\nonumber\\
s_2&=&-f=t_0,\nonumber\\
r_1&=&-v_2.
\eeqy

We emphasize here that these relationships are independent of the masses of the light hadrons, the value of $q^2$, and even
the number of colors in QCD. The only criterion required is that the mass of the heavy quark be much larger than some scale
`$\Lambda_{{\rm QCD}}$', which determines the regime of non-perturbative QCD dynamics.

 In addition to the relationships among the form factors, eqns. (\ref{lorentza}) and (\ref{heavya}) can be used to determine how
the form factors of eqn. (\ref{lorentza}) scale with the mass of the heavy quark or heavy meson. For instance,
\begin{equation}
f_++f_-\simeq \sqrt{m_D},\,\,\,g^\prime\simeq \frac{1}{\sqrt{m_D}},\,\,\,a_++a_-\simeq\frac{1}{\sqrt{m_D^3}}.
\end{equation}

\section{Pole Model}

We assume that the process $D\to
K$ through the current $\bar{s}\Gamma c$ proceeds as a two-step process, in which there is first the strong
`decay' $D\to K D_s^i$, where $D_s^i$ is any of the $D_s$ resonances with the appropriate quantum numbers,
followed by the `leptonic' decay of the $D_s^i$ through the current $\bar{s}\Gamma c$. This means that we
write
\beq
\left<K^{(*)}\left|\bar{s}\Gamma c\right|D^{(*)}\right>=\sum_{i,{\rm spins}}\frac{\left<0\left|\bar{s}\Gamma
c\right|D_s^i\right>\left<D_s^iK^{(*)}|D^{(*)}\right>}{q^2-m_{D_s^i}^2},
\eeq
where the sum includes all possible $D_s^i$ states allowed, and over all spins of those states.

Within HQET we have already discussed, in the previous section, all of the `semileptonic' matrix elements
$\left<K^{(*)}\left|\bar{s}\Gamma c\right|D^{(*)}\right>$. Now we turn to the strong and `leptonic' matrix elements.

For the leptonic processes $\left<0\left|\bar{s}\Gamma c\right|D_s^i\right>$, we find from HQET that these matrix elements can
be written
\beq
\left<0\left|\bar{s}\Gamma c\right|D_s^i(v)\right>={\rm Tr}\left[\Gamma {\cal M}_{D_s^i}(v)\right] f_{D_s^i},
\eeq
where ${\cal M}_{D_s^i}(v)$ is taken to be the Falk representation of the state, and $f_{D_s^i}$ is its decay constant. 
This immediately implies that only states whose Falk representations have no free Lorentz indices can contribute
in the pole model, as only these have non-zero leptonic matrix elements. Thus, only the $(0^-,1^-)$ and $(0^+,1^+)$ $D_s$ 
multiplets are of interest.

For the strong vertices, we use the form developed by the author \cite{roberts}, namely that
\beqy
\left<D_s^i(v)K(p)|D^{(*)}(v)\right>&=&{\rm Tr}\left[a_i \slash{p}\gamma_5\overline{{\cal M}}_{D_s^i}(v){\cal
M}_D^{(*)}(v)\right],\nonumber\\
\left<D_s^i(v)K^*(p,\varepsilon)|D^{(*)}(v)\right>&=&{\rm Tr}\left[\left(b_i\slash{\varepsilon}^*+c_i\varepsilon^*\cdot v\right)
\slash{p}\overline{{\cal M}}_{D_s^i}(v){\cal M}_D^{(*)}(v)\right].
\eeqy
Here $a_i$, $b_i$ and $c_i$ are phenomenological constants that are independent of the mass of the heavy quark in the parent hadron.

There is one subtlety here in that the emission of the light kaon cannot change the velocity of the heavy hadron, so that the
parent $D^{(*)}$ and daughter $D_s^i$ move with the same velocity. However, we will simply take the forms above as being valid for
all kinematics.

With these forms, we can then calculate the semileptonic matrix elements, and extract the contributions to the various form
factors. Doing this, we find, for instance, that
\beqy \label{poleforms}
f_+&=&2\sqrt{\frac{2}{m_D}}\frac{m_D m_{D_s}+q^2}{m_{D_s}^2}\frac{f_{D_s} a_{D_s}}{1-\frac{q^2}{m_{D_s}^2}},\nonumber\\
f_-&=&-2\sqrt{\frac{2}{m_D}}\frac{m_D m_{D_s}^3+m_{D_s}m_D^3-2m_{D_s}^2m_K^2-m_{D_s}m_Dm_K^2+m_{D_s}^2q^2-m_{D_s}m_D q^2+
m_D^2q^2+m_K^2q^2-q^4}{m_{D_s}^4}\frac{f_{D_s} a_{D_s}}{1-\frac{q^2}{m_{D_s}^2}}\nonumber\\
&-&2\sqrt{\frac{2}{m_D}}\frac{m_D^3 m_{D_s^\prime}+m_D^3+m_{D_s^\prime}m_K^2-m_Dm_K^2-m_{D_s^\prime}q^2-m_D q^2}
{m_{D_s^\prime}^3}\frac{f_{D_s^\prime} a_{D_s^\prime}}{1-\frac{q^2}{m_{D_s^\prime}^2}},\nonumber\\
g&=&-2\sqrt{\frac{2}{m_D}}\frac{m_D +m_{D_s}}{m_{D_s}^2}\frac{f_{D_s} c_{D_s}}{1-\frac{q^2}{m_{D_s}^2}},\nonumber\\
f&=&-2\sqrt{\frac{2}{m_D}}\frac{m_{D_s^\prime}m_D^2+m_D^3+m_{D_s^\prime}m_K^2-m_Dm_K^2-m_{D_s^\prime}q^2-m_Dq^2}{m_{D_s}^2}\frac{f_{D_s^\prime} c_{D_s^\prime}}{1-\frac{q^2}{m_{D_s^\prime}^2}},\nonumber\\
a_+&=&2\sqrt{\frac{2}{m_D^3}}\frac{\left[b_{D_s^\prime}(m_D m_{D_s^\prime}+q^2)-m_Dc_{D_s^\prime}(m_{D_s^\prime}+m_D)\right]}{m_{D_s^\prime}^2}\frac{f_{D_s^\prime}}{1-\frac{q^2}{m_{D_s^\prime}^2}},\nonumber\\
a_-&=&-2\sqrt{\frac{2}{m_D^3}}\frac{\left[b_{D_s}(m_D^2 m_{D_s}+m_D^3+m_{D_s}m_K^2-m_Dm_K^2-m_{D_s}q^2-m_D q^2)+2m_Dm_K^2c_{D_s}\right]}
{m_{D_s}^3}\frac{f_{D_s}}{1-\frac{q^2}{m_{D_s}^2}}\nonumber\\
&-&2\sqrt{\frac{2}{m_D^3}}\frac{1}{m_{D_s^\prime}^4}\frac{f_{D_s^\prime}}{1-\frac{q^2}{m_{D_s^\prime}^2}}\left[c_{D_s^\prime}
(2m_{D_s^\prime}m_Dm_K^2-m_{D_s^\prime}^3m_D-m_{D_s^\prime}^2m_D^2)\right.\nonumber\\
&&+\left.b_{D_s^\prime}(m_D m_{D_s^\prime}^3+m_{D_s^\prime}m_D^3-2m_{D_s^\prime}^2m_K^2-m_{D_s^\prime}m_Dm_K^2+
m_{D_s^\prime}^2q^2-m_{D_s^\prime}m_D q^2+m_D^2q^2+m_K^2q^2-q^4)\right].
\eeqy
In these equations, the subscript $D_s$ refers to either member of the $\left(0^-,1^-\right)$ ground state multiplet of $D_s$ states, while $D_s^\prime$
refers to the $\left(0^+,1^+\right)$ multiplet.
 
In their model for the form factors, Bauer, Stech and Wirbel \cite{bsw} write
\beqy \label{bsw}
f_+&=&\frac{h_1}{1-\frac{q^2}{m_{D_s}^2}},\nonumber\\
f_-&=&\frac{m_D^2-m_K^2}{q^2}\left[\frac{h_0}{1-\frac{q^2}{m_{D_s^\prime}^2}}-\frac{h_1}{1-\frac{q^2}{m_{D_s}^2}}\right],\nonumber\\
g&=&\frac{1}{m_D+m_{K^*}}\frac{h_V}{1-\frac{q^2}{m_{D_s}^2}},\nonumber\\
f&=&\left(m_D+m_{K^*}\right)\frac{h_{A_1}}{1-\frac{q^2}{m_{D_s^\prime}^2}},\nonumber\\
a_+&=&\frac{1}{m_D+m_{K^*}}\frac{h_{A_2}}{1-\frac{q^2}{m_{D_s^\prime}^2}},\nonumber\\
a_-&=&\frac{2 m_{K^*}}{q^2}\left[\frac{h_{A_0}}{1-\frac{q^2}{m_{D_s}^2}}-
\frac{m_D+m_{K^*}}{2m_{K^*}}\frac{h_{A_1}}{1-\frac{q^2}{m_{D_s^\prime}^2}}+
\frac{m_D-m_{K^*}}{2m_{K^*}}\frac{h_{A_2}}{1-\frac{q^2}{m_{D_s^\prime}^2}}\right].
\eeqy

The forms written in the last two sets of equations are similar, not unexpectedly, up to the extra kinematic dependence present
in the numerators of eqns. (\ref{poleforms}).

\section{Results and Discussion}

The numerators of the form factor expressions obtained are, in general, $q^2$ dependent, and $q^2$ ranges from
$q^2=m_\ell^2$, where $m_\ell$ is the mass of the `lepton' emitted in the decay, to $q^2=(m_{D^{(*)}}-m_{K^{(*)}})^2$.
In the latter case, $q^2$ scales with the mass of the heavy quark, so it is instructive to examine the relationships
among the form factors in the two extremes, independently. We note, however, that the predictions for the relationships
among the form factors arising from HQET are independent of $q^2$, and so should hold at any value of $q^2$.

\subsection{$q^2=(m_{D^{(*)}}-m_{K^{(*)}})^2$}

Of the numerous relations among the form factors shown in eqn.
(\ref{relationships}), very few are satisfied in the pole model we have
constructed, in the $q^2=(m_{D^{(*)}}-m_{K^{(*)}})^2$ limit, at least at first
sight. If we set $m_c$ to infinity, so that all the heavy mesons are
approximately degenerate, and take the kaon and $K^*$ as approximately massless, we still find
that very few of the relationships are satisfied. However, if we also take into
account the scaling relationships of the form factors, then we find that most of
the relationships do indeed hold. 

For instance, we find that $f_+-f_-=r_2$ so that $f_+-f_--r_2=0$ should hold. 
In fact, in the pole model, the right hand side of this latter equation does not
vanish. However,
\begin{eqnarray}
f_+-f_-&\simeq& \sqrt{m_D},\nonumber\\
r_2&\simeq& \sqrt{m_D},\nonumber
\end{eqnarray}
while, in the pole model
\begin{eqnarray}
f_+-f_--r_2\simeq \frac{1}{\sqrt{m_D}},\nonumber
\end{eqnarray}
meaning that `at leading order' for these form factors, the relationship
$f_+-f_-=r_2$ is satisfied. This is the manner in which all of the form factor
relationships that do hold, become satisfied; the observed violations are of higher order in $1/m_c$.

The form factors, or combinations of form factors, that vanish exactly, namely
\beq 
a_+^\prime-a_-^\prime=h^\prime=h_1=t_3=t_+^\prime-t_-^\prime=v_+-v_-=0
\eeq
are not satisfied in the pole model, at the $q^2=(m_{D^{(*)}}-m_{K^{(*)}})^2$ limit. Setting all heavy masses to be the same, and
the kaon ($K^*$) mass to be zero does not help. In fact, these form factor combinations are all proportional to $1/m_{K^{(*)}}$, in this limit.
Thus, it is not clear how else to interpret these results but to say that these relationships are not satisfied in this pole
model. Furthermore, use of scaling arguments may work for the combinations $a_+^\prime-a_-^\prime$, $t_+^\prime-t_-^\prime$ and 
$v_+-v_-$, but there are no clear scaling arguments that can be made to say that the non-vanishing contributions to $h^\prime$, $h_1$ and $t_3$
are in fact consistent with zero, at leading order. We note that using such scaling arguments suggests that while the combinations 
$t_+^\prime-t_-^\prime$ and $v_+-v_-$ do indeed vanish at leading order (modulo a factor of $1/m_{K^{(*)}}$), 
the combination $a_+^\prime-a_-^\prime$ does not.

\subsection{$q^2=0$}

In this limit, fewer of the relationships among the form factors are satisfied, even when the scaling arguments are used. For
instance, 
\begin{eqnarray}
f_++f_-&\simeq& \frac{1}{\sqrt{m_D}},\nonumber\\
g_+^\prime+g_-^\prime&\simeq& \frac{1}{\sqrt{m_D}},\nonumber
\end{eqnarray}
and
\begin{eqnarray}
f_++f_--(g_+^\prime+g_-^\prime)\simeq \frac{1}{\sqrt{m_D}}.\nonumber
\end{eqnarray}
In the $q^2=(m_{D^{(*)}}-m_{K^{(*)}})^2$ limit, this difference scales like $1/\sqrt{m_D^3}$, thus vanishing at leading order. Here,
in the $q^2=0$ limit, this difference clearly does not vanish, not even when scaling arguments are invoked.

\section{Conclusions}

In the pole model that we have constructed, a number of the HQET-predicted relationships among form factors are found to be violated at all values of
$q^2$. Grinstein \cite{grinstein} has argued that in the combined heavy quark, chiral and large $N_c$ limits, the few relationships examined for pseudoscalar meson
transitions ($B \to \pi$), in a pole model similar to the one constructed herein, do hold. However, the relationships shown in eqn.
(\ref{relationships}) are obtained only in the limit of a heavy quark, and are `unaware' of the mass of the light mesons and the number of colors.
They should therefore hold without the need to invoke these additional limits. 

In view of our results, we suggest that the available data on $D$ semileptonic decays be reanalyzed using a paramterization of the form factors other
than the monopole form. Certainly, at least the form factor $f$ should be allowed to depart from the `simple' monopole form. Furthermore, if funding for CLEO-c 
is approved, analysis of new data obtained on $D$ semileptonic decays should relax the `monopole assumption' for the form factors. Because of the limited phase
space, the form of the parametrization used will probably not have important consequences for $D$ meson decays, not even for $D\to\pi$ processes.
However, the differences in parametrization can lead to very different results for $B$ decays, and are therefore of crucial importance for the
extraction of $V_{ub}$, for instance. This becomes even more important if the form factors for $D\to\pi$ (or $D\to\rho$) semileptonic decays, in
conjunction with the scaling predictions of HQET, are used in analysis of $B\to\pi$ (or $B\to\rho$) processes.  \\\\

The author thanks J. L. Goity for reading the manuscript, and for discussions.
This work was
supported by the National Science Foundation through grants \# PHY-9457892 and \# PHY 9820458. This work was also supported 
by the Department of Energy through contract DE-AC05-84ER40150, under which the Southeastern Universities Research Association (SURA) operates the Thomas
Jefferson National Accelerator Facility (TJNAF), and through contract DE-FG05-94ER40832.

\end{document}